# A Single Atlas is All You Need: Decoder-Side Gaussian Splatting for Immersive Video

Dawid Mieloch, *Senior Member, IEEE*, Stuart Perry, *Senior Member, IEEE*

*Abstract*—**Immersive video delivery is bottlenecked by pixel-rate constraints, making the transmission of high-resolution depth maps or explicit 3D volumetric data expensive. Decoder-Side Depth Estimation (DSDE) shifts depth computation to the client, but struggles with complex geometries, inter-view flickering, and non-Lambertian reflections. Conversely, 3D Gaussian Splatting (3DGS) offers state-of-the-art view synthesis, but transmitting splats (or their projected 2D maps) incurs prohibitive bandwidth costs and is poorly aligned with standard video codecs. We propose Decoder-Side Gaussian Splatting (DSGS), a framework that natively replaces the depth-estimation stage of DSDE with feed-forward 3DGS inference, optimizing volumetric scenes entirely on the decoder side from compressed textures and metadata. A central, counterintuitive finding is that lossy compression acts as an implicit low-pass filter stabilizing feed-forward splat prediction: compressed bitstreams exceed lossless quality while shrinking tenfold. Under extreme view sparsity (one 2D atlas comprising 4 input views), DSGS achieves a +5.79 dB BD-PSNR and +0.054 BD-SSIM gain over the DSDE anchor while reducing maximum inter-view ΔIV-PSNR from 17.2 dB to 6.4 dB, minimizing the domain shift between transmitted and virtual viewports.**

## I. Introduction

To provide for the possibility of high-quality virtual navigation for the end user of virtual reality systems, standard immersive video transmission techniques require multiple views and depth maps, quickly exceeding hardware decoder limits [1]. Decoder-side depth estimation (DSDE [2]) addressed this by shifting depth (geometry) calculation to the client, thereby allocating all available bitrate to textures. Therefore, in the DSDE framework, the encoder completely omits explicit geometry (depth maps) and transmits only sparse input views as 2D atlases, along with camera metadata. The client device is then responsible for estimating the missing geometry locally, using a depth estimator to gather data required to render novel views. Unfortunately, pixel- and segment-wise depth estimators (like MPEG Immersive Video Depth Estimation [2]) fail to estimate reliable geometry in real time, making the usefulness of DSDE-based compression in real-world applications relatively low. Subsequent refinements of DSDE, such as Input Depth Map Assistance [3], reduced runtime by transmitting partial geometry hints. However, depth maps are not suitable for handling non-Lambertian surfaces [4], leading to structural degradation and a noticeable visual gap between synthesized and transmitted views.

This work was supported by the Ministry of Science and Higher Education of the Republic of Poland.

Dawid Mieloch is with the Institute of Multimedia Telecommunications, Poznan University of Technology, 60-236 Poznan, Poland (e-mail: dawid.mieloch@put.poznan.pl).

Stuart Perry is with the Perceptual Imaging Laboratory (PILab), School of Electrical and Data Engineering, University of Technology Sydney, Ultimo, NSW, Australia (e-mail: stuart.perry@uts.edu.au).

3D Gaussian Splatting (3DGS [5], [6]) has already established itself as a versatile, widely used representation for immersive media [7], [8]. With highly improved handling of complex, non-Lambertian objects, and fast, high-quality synthesis of virtual views, 3DGS has become the new state-of-the-art in immersive video applications. As with any emerging video-related technology, to ensure interoperability and performance on resource-constrained devices, the Moving Picture Experts Group (MPEG) is actively standardizing Gaussian Splat Coding (GSC) [9]. A key part of this initiative is the fast-track amendment to the Video-based Point Cloud Compression (V-PCC) standard. By projecting splat attributes into 2D maps, the framework prioritizes rapid market adoption by exploiting universally available, hardware-accelerated 2D video codecs to compress these maps, bypassing the need for specialized geometry-decoding hardware. Therefore, a primary challenge in 3DGS compression is the efficient representation of the data attributes fed into the video.

Unfortunately, current video-based GS compression methods (such as LGSCV [10], GIFStream [8], and CompSplat [11]) often rely on computationally expensive techniques, such as Parallel Linear Assignment Sorting (PLAS) or Morton scans, to convert 3DGS into 2D maps, limiting their applicability on lightweight devices, even with advanced strategies for efficient adaptive streaming and pruning of the Gaussian primitives [12]. Moreover, representing splat attributes (such as covariance matrices or their positions) as a video channel makes it very hard to compress them efficiently using traditional video compression methods. Standard video codecs are heavily optimized for the human visual system, preserving psychovisual similarity while discarding high-frequency details. Applying such compression to abstract non-visual data structures leads to unpredictable data degradation and severe geometric and visual artifacts during rendering. Other efforts in 3DGS compression [13], focus on progressive, rate-distortion-optimized compression of color and semantic attributes using Implicit Neural Representation (INR) based hyperpriors to model attribute distributions. While effective, these methods still require the transmission of quantized splat parameters.

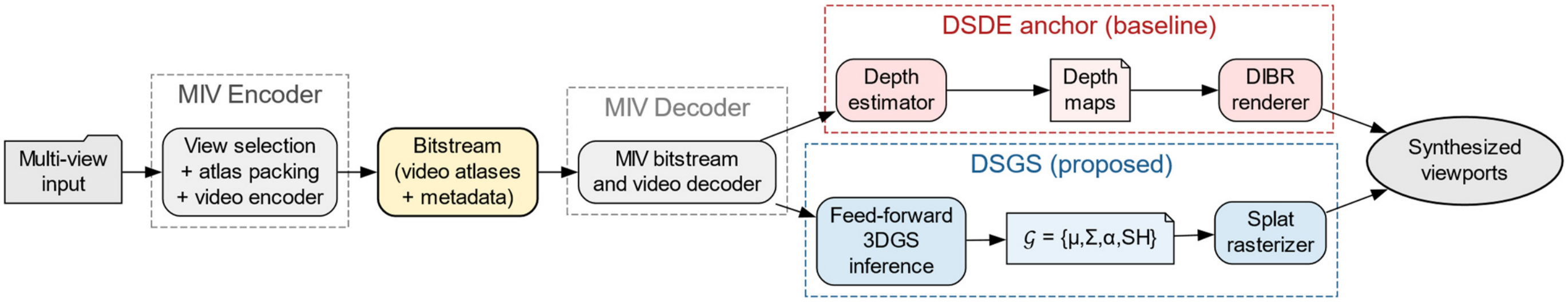

Fig. 1. Comparison of MIV decoder-side processing in DSDE (baseline) and DSGS (proposed). Both pipelines consume the same MIV bitstream; DSGS replaces the IVDE → depth maps → DIBR module sequence with a feed-forward 3DGS predictor and splat rasterizer.

To overcome the high computational cost of per-scene 3DGS optimization, a growing line of work predicts Gaussian primitives directly from sparse views in a single forward pass. Splatter Image initiated the paradigm [14] for single-view object reconstruction and, with pixelSplat [15], for binocular scene reconstruction with probabilistic depth sampling. It was later scaled to multiview inputs by MVSplat [16], which builds plane-sweep cost volumes (analogous in spirit to traditional DSDE) to localize Gaussian centers. DepthSplat [17] explicitly couples feed-forward 3DGS with multiview depth estimation, demonstrating that strong depth priors and Gaussian rendering are mutually reinforcing. More recent pose-free variants such as NoPoSplat [18], PF3plat [19], FLARE [20], and AnySplat [21] further demonstrate that high-quality 3D geometry can be inferred from sparse views in under 0.5 s on commercial hardware, proving the feasibility of fully client-side inference. Recent advances in on-device 3DGS training [22] confirm the feasibility of client-side execution on mobile hardware.

To merge the advantages of DSDE and 3DGS, our proposed **Decoder-Side Gaussian Splatting (DSGS) approach for immersive compression, replaces the flawed depth estimation step of DSDE with a direct 3DGS optimization pipeline (Fig. 1).** Therefore, instead of compressing and transmitting explicit or projected splat attributes, we propose leveraging the existing MPEG Immersive Video (MIV) DSDE framework to transmit only compressed sparse views.

The 2D atlases of MIV DSDE, which contain a subset of views from a multiview sequence, are used to initialize a 3DGS model on the client side, effectively rendering both explicit geometry transmission and estimation redundant. As available methods can provide high-quality results in a feed-forward manner, the most computationally costly step of training new model is not required, making the rendering of new virtual views possible in a close-to-real-time way. Moreover, transmission of atlases is more effective than transmission of 3DGS attributes, as any state-of-the-art video compression method can be used to encode and decode them effectively.

A concurrent learned-codec approach to decoder-side feed-forward 3DGS has been explored by Yang and Qin [7] for stereo human-centric content with jointly trained semantic autoencoders. DSGS instead operates on already standardized, unmodified MIV DSDE bitstream, making it useful for any multiview sequence. A complementary line of work [23] uses 3DGS as a geometric prior to guide 2D multiview compression, whereas DSGS inverts this relationship: it consumes 2D atlases to produce 3DGS at the decoder.

## II. Decoder-side Gaussian Splatting

The contribution of DSGS is not the use of a feed-forward 3DGS model, as any of the recent predictors [14]-[21], [24] could be used in the pipeline. Rather, the contribution lies in three observations that, together, redefine the role of decoder-side processing in immersive video:

(i) **Geometry transmission becomes redundant, as depth estimation is replaced by direct volumetric inference.** The MIV bitstream already carries everything a feed-forward splat predictor requires, i.e., non-pruned texture atlases and exact camera metadata, so no syntax extension is necessary. DSGS is therefore deployable within unmodified MIV DSDE encoders, in contrast to upcoming V-PCC-based GSC profiles [9] that require new sub-bitstreams for projected splat attributes.

(ii) **Cost-volume depth estimation and feed-forward 3DGS share the same input but differ fundamentally in output regularity.** Both IVDE and MVSplat-style predictors [16] consume multiview images. Still, the former emits per-pixel scalar depth (sensitive to view-dependent effects and disparity quantization). In contrast, the latter emits a continuous volumetric field whose covariance and opacity attributes directly absorb non-Lambertian phenomena.

(iii) **The encoder-decoder asymmetry of MIV synergizes with the noise sensitivity of feed-forward 3DGS**. As detailed in Section III-B, lossy 2D codec quantization acts as an implicit low-pass filter on the input to the splat predictor, improving synthesis quality rather than degrading it – a property without analog in the DSDE anchor, where compression monotonically degrades depth quality.

The remainder of this section describes the concrete instantiation of this paradigm. In our proposal, the MIV encoder architecture is unchanged from the standard MIV DSDE profile. A subset of input views of a compressed multiview sequence is selected, packed into 2D video atlases, and transmitted in a single bitstream alongside camera extrinsic and intrinsic parameters.

Then, the decoded texture atlases and metadata bypass traditional cost-volume depth estimation. Formally, instead of mapping decoded 2D textures $\mathcal{I}$ and camera poses $\mathcal{P}$ to depth maps, the proposed framework models a direct mapping function $\mathcal{F}: \mathcal{I}, \mathcal{P} \rightarrow \mathcal{G}$, outputting the explicit 3DGS state space $\mathcal{G} = \{\mu, \Sigma, \alpha, SH\}$ (center $\mu$, covariance matrix $\Sigma$, opacity $\alpha$, and spherical harmonics $SH$). By inferring view-dependent color attributes ($SH$) locally from the decoded textures, DSGS entirely circumvents the severe quantization and dimensionality reduction required in standard GSC transmission.

To overcome the prohibitive latency introduced by standard Stochastic Gradient Descent (SGD) optimization required for training a new model, our framework integrates a feed-forward recurrent Gaussian splatting model at the decoder. Traditional depth estimators rely on stereoscopic cost-volume matching, which typically scales at $\mathcal{O}(N^2)$ depending on the number of disparity search steps and spatial resolution. In contrast, the feed-forward inference of the DSGS model executes in $\mathcal{O}(N)$ time relative to depth complexity, bounding the volumetric generation cost to near real time on client-side GPUs or NPUs.

While any feed-forward 3DGS method can be used in the proposed framework, the described implementation used the ReSplat method [24]. It utilizes the rendering error as a feedback signal to iteratively refine the 3D Gaussians without explicitly computing gradients. Furthermore, it initializes a compact reconstruction model in a subsampled space, producing fewer Gaussians compared to per-pixel representations, enabling faster tile-based rasterization and rendering. Contribution [24] provides further comparative analysis against representative feed-forward predictors, including DepthSplat [17], demonstrating that ReSplat achieves state-of-the-art quality at a lower computational cost. The work presented in this contribution inherits that benchmark rather than reproducing it, and focuses on the orthogonal question of integrating feed-forward 3DGS into a standardized immersive-video coding pipeline.

While recent advancements in feed-forward 3DGS have introduced highly capable pose-free architecture, these models dedicate computational resources to estimating camera parameters from uncalibrated images. In the strict context of the MIV standard, exact camera poses and metadata are explicitly transmitted alongside the visual data as a mandatory sub-bitstream. Therefore, utilizing pose-free models would introduce unnecessary computational redundancy. Accordingly, to bridge the standardized MIV framework with the most widely used representation of camera parameters in 3DGS methods, the proposal converts MIV camera metadata (intrinsics and Euler extrinsics) to the OpenCV format.

The dynamic initialization of 3D Gaussians is highly sensitive to high-frequency noise and micro-textures, leading the optimization process to generate redundant, high-variance splats in space [25], [26]. By using lossy 2D video compression, the encoded atlases effectively undergo low-pass filtering [27]. This implicit denoising forces the feed-forward network to focus on the dominant, coherent geometry, thereby stabilizing the covariance matrix $\Sigma$ and reducing the variance of generated attributes. While explicit geometric regularization techniques have been proposed to stabilize 3DGS from sparse inputs [28], our framework achieves a similar effect natively through the inherent low-pass filtering properties of lossy 2D video compression. Furthermore, based on the information bottleneck principle used in recent scalable feed-forward 3DGS architectures (e.g., ZPressor [29]), limiting the input information via compression forces the network to discard redundancy, generalize better, and yield stable 3D geometry.

## III. Experimental Results

The proposed framework was evaluated using the methodology described in the MIV Common Test Conditions (CTC [30]), with two changes: the number of atlases was reduced to one (max 4 views) or two (max 8 views), and all views were scaled to 960×540 resolution to match the internal resolution of the ReSplat pre-trained model. The experiments used an All-Intra configuration, and video compression was performed by VVenc [31].

The anchor across all experiments is the MIV DSDE profile, using the reference IVDE depth estimator [2]. The choice of IVDE as the single DSDE depth-estimator anchor is justified by [2], which provides a comprehensive comparison of IVDE against alternative state-of-the-art depth estimators (including learning-based methods) within the MIV DSDE framework and establishes IVDE as the reference baseline; we therefore inherit this benchmark without repeating cross-estimator comparison.

The integrated feed-forward pipeline was evaluated across MPEG Common Test Conditions sequences at five distinct rate points: RP0 (lossless) - RP4 (highest compression). The dataset includes 14 multiview test sequences spanning mandatory and non-mandatory content (Classes J, W, D, E, L), incorporating challenging indoor and outdoor rigs. Computer-generated omnidirectional sequences (Classes A, B, C) were excluded, as current feed-forward network architectures assume pinhole projections and can struggle with equirectangular distortions.

In the MIV CTC methodology, quality is measured by comparing the views rendered by the decoder at the exact positions of all views in a multiview sequence to their uncompressed counterparts. Quality is assessed using Bjøntegaard Delta PSNR (BD-PSNR over PSNR and IV-PSNR [32]) and BD-SSIM over IV-SSIM (Immersive Video Structural Similarity [33]) to evaluate structural integrity in 6DoF environments. These metrics were also utilized to measure inter-view consistency, i.e., the maximum difference (delta) between qualities of all rendered views.

### A. *Results*

The evaluation primarily focuses on the extreme view sparsity scenario (1 atlas, comprising 4 physical views), aligning with the premise of pushing decoder-side synthesis to its limits. Tests under denser setups (e.g., 2 atlases with 8 views) indicated that both DSGS and the DSDE anchor perform comparably (yielding marginal average differences of approximately +0.1 dB in BD-PSNR in favor of DSGS), the severe limitations of traditional depth estimation (and conversely, the distinct advantages of our direct 3DGS optimization) become clearly evident only when the availability of visual data is heavily constrained.

As shown in Table I, the average objective gain over DSDE is +5.79 dB IV-PSNR, but the critical advantage of this approach lies in inter-view consistency. Table II illustrates the maximum Delta IV-PSNR, which quantifies the quality gap between the best and worst synthesized views. The traditional DSDE anchor exhibits an average maximum Delta IV PSNR of 17.2 dB, leading to severe visual flickering as the user navigates between virtual and close-to-physical viewpoints. Utilizing

DSGS reduces this metric to an average of 6.4 dB. This reduction demonstrates that rendering all views natively from the splatting engine effectively minimizes the domain shift, facilitating a more seamless immersive experience.

TABLE I
BD-PSNR AND BD-SSIM OF DSGS OVER DSDE ANCHOR

| Sequence | | BD-PSNR Y-PSNR [dB] | BD-PSNR IV-PSNR [dB] | BD-SSIM IV-SSIM |
|---|---|---|---|---|
| Kitchen | J01 | 10.96 | 11.12 | 0.1355 |
| Cadillac | J02 | 9.05 | 10.99 | 0.0857 |
| Mirror | J03 | 8.85 | 8.74 | 0.1147 |
| Fan | J04 | 7.44 | 9.82 | 0.0896 |
| Group | W01 | 6.37 | 6.73 | 0.0820 |
| Dancing | W02 | 8.61 | 9.34 | 0.1268 |
| Painter | D01 | 0.92 | 3.98 | 0.0047 |
| Breakfast | D02 | 7.35 | 8.52 | 0.0756 |
| Barn | D03 | 3.84 | 3.80 | 0.0174 |
| Frog | E01 | 11.58 | 15.35 | 0.1487 |
| Carpark | E02 | -2.27 | -2.23 | -0.0105 |
| Street | E03 | -6.14 | -6.22 | -0.0097 |
| Fencing | L01 | -0.14 | 0.56 | -0.0037 |
| CBABasketball | L02 | 2.67 | 3.95 | 0.0249 |
| MartialArts | L03 | 8.82 | 12.87 | 0.1403 |
| **Average** | | **3.78** | **5.79** | **0.0538** |

TABLE II
MAXIMUM INTER-VIEW QUALITY DIFFERENCES (Δ) ACROSS SYNTHESIZED VIEWS: DSDE ANCHOR VS. DSGS

| Sequence | | ΔY-PSNR [dB] DSDE | ΔY-PSNR [dB] DSGS | ΔIV-PSNR [dB] DSDE | ΔIV-PSNR [dB] DSGS | ΔIV-SSIM DSDE | ΔIV-SSIM DSGS |
|---|---|---|---|---|---|---|---|
| Kitchen | J01 | 21.89 | 7.61 | 23.81 | 6.05 | 0.25 | 0.02 |
| Cadillac | J02 | 22.06 | 7.21 | 26.54 | 9.32 | 0.21 | 0.04 |
| Mirror | J03 | 21.89 | 6.82 | 23.76 | 6.55 | 0.23 | 0.04 |
| Fan | J04 | 20.55 | 6.10 | 23.53 | 6.57 | 0.18 | 0.03 |
| Group | W01 | 32.40 | 8.81 | 32.74 | 9.76 | 0.27 | 0.10 |
| Dancing | W02 | 25.48 | 8.96 | 26.21 | 9.85 | 0.24 | 0.05 |
| Painter | D01 | 7.23 | 1.45 | 7.24 | 2.69 | 0.03 | 0.02 |
| Breakfast | D02 | 8.64 | 7.77 | 9.18 | 7.50 | 0.21 | 0.09 |
| Barn | D03 | 7.80 | 8.55 | 8.05 | 8.46 | 0.18 | 0.13 |
| Frog | E01 | 7.84 | 3.80 | 7.68 | 3.06 | 0.22 | 0.02 |
| Carpark | E02 | 9.88 | 1.03 | 10.51 | 1.55 | 0.01 | 0.01 |
| Street | E03 | 10.88 | 0.52 | 10.00 | 0.41 | 0.01 | 0.00 |
| Fencing | L01 | 16.76 | 4.91 | 16.59 | 6.46 | 0.03 | 0.02 |
| CBABasketball | L02 | 18.52 | 6.07 | 18.80 | 7.32 | 0.15 | 0.10 |
| MartialArts | L03 | 13.61 | 9.28 | 13.31 | 10.37 | 0.23 | 0.04 |
| **Average** | | **16.36** | **5.93** | **17.20** | **6.40** | **0.16** | **0.05** |

DSGS underperforms the DSDE anchor on three sequences: E02, E03, and L01. These are precisely the sequences with the smallest source-view count in the dataset (9 cameras). The relative sparsity of 4 out of 9 views is lower than for the remaining sequences, where 4 views are selected from larger pools. Two consequences follow. First, the DSDE benefits from narrower baselines between transmitted views, thereby improving the reliability of cost-volume matching, i.e., the operating points where traditional depth estimation is least challenged. Second, the MIV CTC evaluation averages quality across all original camera positions, so with only 9 evaluation viewpoints, a small number of difficult synthesis positions disproportionately affects the per-sequence mean. Negative results, therefore, mark an area where the advantage of DSGS diminishes, rather than indicate a failure of the framework.

The runtime of the decoder dropped significantly from (on average) more than 15 minutes (AMD Ryzen 9 CPU) to almost real-time, as the feed-forward inference took less than 1 second, with further rendering performed with more than 500 fps (Nvidia H100 GPU). Peak GPU memory use was <10 GB.

## *B. Impact of input compression on volumetric synthesis*

Conventional rate–distortion analysis assumes that synthesis quality decreases monotonically with input bitrate. DSGS violates this assumption, with direct consequences for the design of asymmetric Rate-Distortion-Complexity Optimization (RDCO). Applying compression to the 2D atlases yielded higher scores than the uncompressed baselines (RP0). Table III shows that the lossless RP0 atlas is not the optimal operating point for feed-forward synthesis: transitioning to lossy RP1 yields an around tenfold bitrate reduction and an IV-SSIM increase from 0.925 to 0.927, with PSNR and IV-PSNR also peaking at RP1–RP2. Quality degrades only beyond RP3.

TABLE III
AVERAGE BITSTREAM SIZE AND OBJECTIVE QUALITY FOR DSDE AND DSGS: GREEN – BEST RESULT IN COLUMN FOR GIVEN NUMBER OF ATLASES, YELLOW – SECOND BEST RESULT

| Num of atlases | Rate point | Average bitstream size [kB] | Average PSNR [dB] DSDE | Average PSNR [dB] DSGS | Average IV-PSNR [dB] DSDE | Average IV-PSNR [dB] DSGS | Average IV-SSIM DSDE | Average IV-SSIM DSGS |
|---|---|---|---|---|---|---|---|---|
| 1 | RP0 | 3107 | 17.87 | 24.20 | 24.09 | 31.58 | 0.860 | 0.925 |
| | RP1 | 297 | 17.88 | 24.41 | 24.08 | 31.77 | 0.858 | 0.927 |
| | RP2 | 184 | 17.86 | 24.43 | 24.05 | 31.69 | 0.856 | 0.926 |
| | RP3 | 111 | 17.84 | 24.36 | 23.98 | 31.45 | 0.852 | 0.921 |
| | RP4 | 59 | 17.82 | 23.97 | 23.87 | 30.66 | 0.843 | 0.907 |
| 2 | RP0 | 6090 | 24.90 | 25.19 | 30.90 | 32.59 | 0.947 | 0.931 |
| | RP1 | 570 | 24.87 | 25.45 | 30.89 | 32.87 | 0.945 | 0.932 |
| | RP2 | 352 | 24.78 | 25.55 | 30.73 | 32.91 | 0.942 | 0.932 |
| | RP3 | 213 | 24.59 | 25.53 | 30.39 | 32.70 | 0.936 | 0.928 |
| | RP4 | 113 | 24.05 | 24.91 | 29.78 | 31.66 | 0.923 | 0.912 |

We attribute this to the well-documented sensitivity of feed-forward 3DGS to high-frequency input noise [25], [26]: 2D video codecs often discard the high-frequency micro-textures that the predictor would otherwise misinterpret as geometric structure, producing "floater" splats. In the presented framework, compression seems to make it easier for pre-trained models to estimate 3DGS. Crucially, this phenomenon is practically absent in the traditional DSDE approach, where increased compression is associated with a steady decline in structural similarity (e.g., IV-SSIM drops continuously from 0.860 at RP0 to 0.843 at RP4). This indicates that direct 3DGS optimization is synergistic with the decoder-side applications, turning standard lossy video compression into an ally rather than an obstacle.

Table III also shows that DSDE requires two atlases to reach acceptable quality and exhibits a steep dependence on atlas count. Doubling from one to two atlases at RP1 raises IV-PSNR by 6.81 dB (24.08 to 30.89 dB) and IV-SSIM by 0.087 (0.858 to 0.945), showing that a single atlas is structurally insufficient for reliable cost-volume depth estimation. In contrast, DSGS with a single atlas at RP1 (297 kB, 31.77 dB IV-PSNR) outperforms DSDE with two atlases at RP1 (570 kB, 30.89 dB IV-PSNR) in objective quality while halving the bitstream. This advantage is visually confirmed in Fig. 2. Single-atlas DSDE exhibits severe ghosting and structural breakdown, two-atlas DSDE restores acceptable quality, and DSGS matches or surpasses two-atlas DSDE while consuming half the bitstream. This validates the central premise that, under feed-forward Decoder-Side Gaussian Splatting, *a single atlas is all you need*.

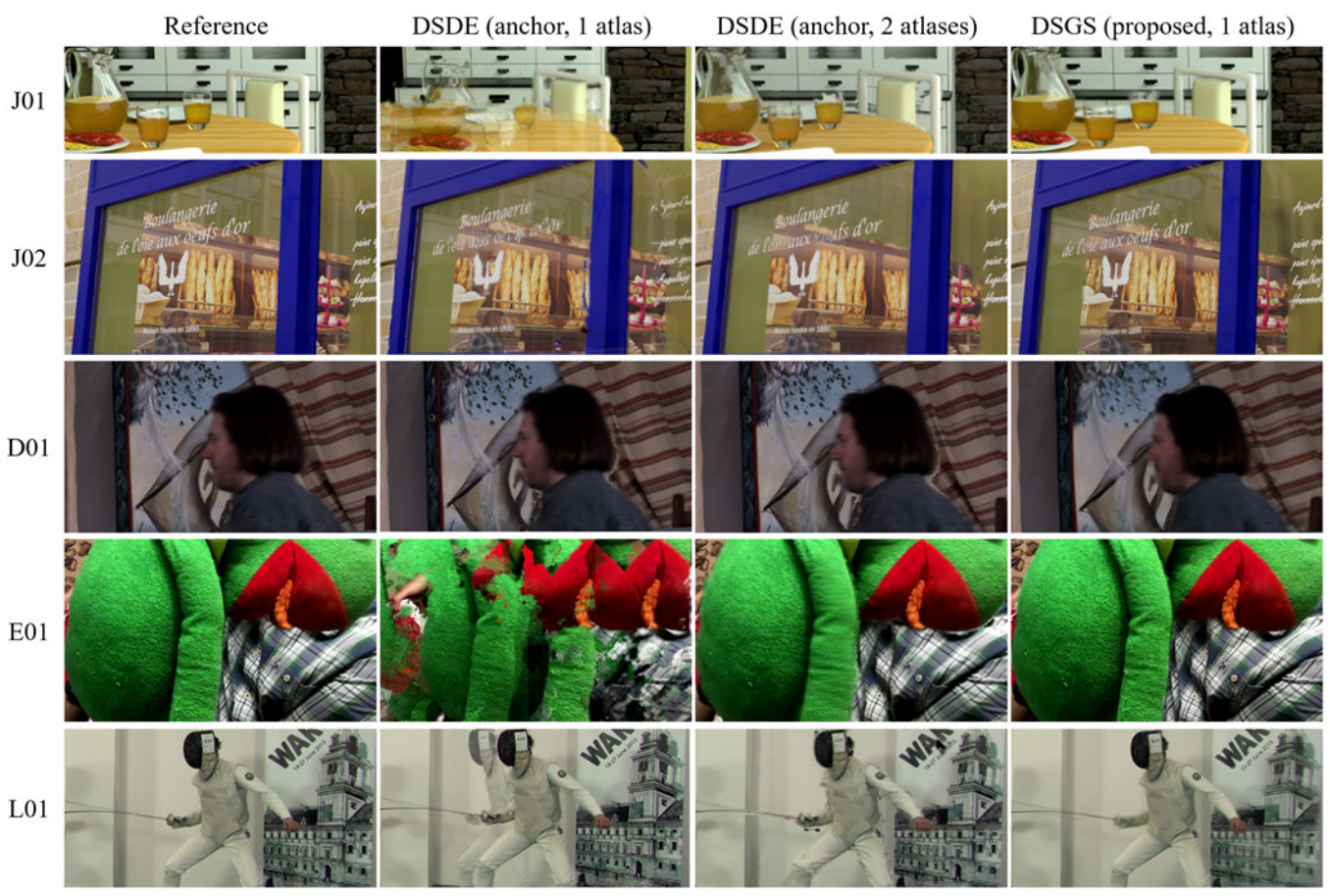

Fig. 2. Visual comparison at RP1. DSDE with a single atlas exhibits severe structural artifacts – ghosting, fragmentation, and duplicated objects. Using two atlases restores acceptable quality. Proposed DSGS achieves comparable or superior quality to two-atlas DSDE using half the bitstream and single atlas.

## IV. Conclusions and Future Work

The proposed DSGS redefines decoder-side processing for immersive video. It outperforms depth estimation under strict bandwidth constraints in final quality and solves the inter-view consistency problem inherent to traditional DIBR rendering. However, the most consequential finding is that standard lossy 2D video compression, long treated as a bandwidth–quality trade-off, here acts as a regularizer that improves 3D synthesis. This property is visible only in the feed-forward decoder-side paradigm and absent from the DSDE pipeline, suggesting that immersive-video standards should be co-designed with the noise sensitivity of their generative back-ends in mind.

A current limitation of applying static feed-forward 3DGS models to sequential MIV data is the lack of temporal consistency. Processing frames independently can lead to minor fluctuations in predicted Gaussian attributes, manifesting as inter-frame flickering or temporal jitter. Future extensions of this framework should incorporate 4D dynamic modeling paradigms. Approaches such as GIFStream [8], which uses time-dependent feature streams to anchor Gaussian primitives across frames, or D-FCGS [34], which introduces control-point-guided motion compensation, offer promising pathways to ensure temporal stability in decoder-side-generated volumes. Future work will also explore extracting lightweight structural features (e.g., sparse point clouds via the Geometry Assistance SEI [2]) and integrating edge-assisted continuous model refinement to guide client-side splatting.

## References

[1] J. M. Boyce et al., "MPEG immersive video coding standard," Proc. IEEE, vol. 109, no. 9, pp. 1521–1536, Sep. 2021.

[2] D. Mieloch et al., "Overview and efficiency of decoder-side depth estimation in MPEG immersive video," IEEE Trans. Circuits Syst. Video Technol., vol. 32, no. 9, pp. 6360–6374, Sep. 2022.

[3] D. Mieloch et al., "A new approach to decoder-side depth estimation in immersive video transmission," IEEE Trans. Broadcast., vol. 69, no. 4, pp. 951–965, Dec. 2023.

[4] S. Pyykölä, N. Joswig and L. Ruotsalainen, "Non-Lambertian Surfaces and Their Challenges for Visual SLAM," IEEE Open Journal of the Computer Society, vol. 5, pp. 430-445, 2024.

[5] B. Kerbl et al. "3D Gaussian splatting for real-time radiance field rendering," ACM Trans. Graph., vol. 42, no. 4, pp. 1–14, Aug. 2023.

[6] Y. Bao et al., "3D Gaussian splatting: Survey, technologies, challenges, and opportunities," IEEE Trans. Circuits Syst. Video Technol., vol. 35, no. 7, pp. 6832–6852, Jul. 2025.

[7] D. Yang et al., "Generalizable 3D Gaussian splatting enabled semantic coding for real-time immersive video communications," arXiv:2604.25330, 2026.

[8] H. Li et al., "GIFStream: 4D Gaussian-based immersive video with feature stream," in Proc. IEEE/CVF Conf. Comput. Vis. Pattern Recognit. (CVPR), Jun. 2025, pp. 21761–21770.

[9] ISO/IEC JTC1/SC29, "Text of ISO/IEC 23090-5 DAM V-PCC for gaussian splats coding," MPEG document N01453, 2026.

[10] Q. Yang, M. Liu, and Y. Xu, "Lightweight 3D Gaussian splatting compression via video codec," arXiv:2512.11186, 2025.

[11] H. Song et al., "CompSplat: Compression-aware 3D Gaussian splatting for real-world video," arXiv:2602.09816, 2026.

[12] Y. Wang et al., "On the efficient adaptive streaming of 3D Gaussian splatting over dynamic networks," IEEE Trans. Circuits Syst. Video Technol., vol. 36, no. 4, pp. 4594–4608, Apr. 2026.

[13] Y.-J. Tseng et al., "CSGaussian: Progressive rate-distortion compression and segmentation for 3D Gaussian splatting," in Proc. IEEE/CVF Winter Conf. Appl. Comput. Vis. (WACV), 2026.

[14] S. Szymanowicz, C. Rupprecht, and A. Vedaldi, "Splatter image: Ultra-fast single-view 3D reconstruction," in Proc. IEEE/CVF Conf. Comput. Vis. Pattern Recognit. (CVPR), Jun. 2024, pp. 10208–10217.

[15] D. Charatan et al., "pixelSplat: 3D Gaussian splats from image pairs for scalable generalizable 3D reconstruction," in Proc. IEEE/CVF Conf. Comput. Vis. Pattern Recognit. (CVPR), Jun. 2024, pp. 19457–19467.

[16] Y. Chen et al., "MVSplat: Efficient 3D Gaussian splatting from sparse multi-view images," in Proc. Eur. Conf. Comput. Vis. (ECCV), 2024.

[17] H. Xu et al., "DepthSplat: Connecting Gaussian splatting and depth," in Proc. IEEE/CVF Conf. Comput. Vis. Pattern Recognit. (CVPR), 2025.

[18] B. Ye et al., "No pose, no problem: Surprisingly simple 3D Gaussian splats from sparse unposed images," in Proc. Int. Conf. Learn. Represent. (ICLR), 2025.

[19] S. Hong et al., "PF3plat: Pose-free feed-forward 3D Gaussian splatting for novel view synthesis," in Proc. Int. Conf. Mach. Learn. (ICML), 2025.

[20] S. Zhang et al., "FLARE: Feed-forward geometry, appearance and camera estimation from uncalibrated sparse views," in Proc. IEEE/CVF Conf. Comput. Vis. Pattern Recognit. (CVPR), Jun. 2025, pp. 21936–21947.

[21] L. Jiang et al., "AnySplat: Feed-forward 3D Gaussian splatting from unconstrained views," ACM Trans. Graphics, vol. 44, no. 6, Dec. 2025.

[22] W. Guo et al., "PocketGS: On-device training of 3D Gaussian splatting for high perceptual modeling," arXiv:2601.17354, 2026.

[23] Y. Huang et al., "3D-LMVIC: Learning-based multi-view image coding with 3D Gaussian geometric priors," arXiv:2409.04013, 2024.

[24] H. Xu, D. Barath, A. Geiger, and M. Pollefeys, "ReSplat: Learning recurrent Gaussian splats," arXiv:2510.08575, 2025.

[25] B. Mildenhall et al., "NeRF in the dark: High dynamic range view synthesis from noisy raw images," in Proc. IEEE/CVF Conf. Comput. Vis. Pattern Recognit. (CVPR), Jun. 2022, pp. 16190–16199.

[26] F. Jiang, Z. Li, and Y. Zhang, "DenoiseSplat: Feed-forward Gaussian splatting for noisy 3D scene reconstruction," arXiv:2603.09291, 2026.

[27] O. K. Al-Shaykh, R. Mersereau, "Lossy compression of noisy images," IEEE Trans. Image Process., vol. 7, no. 12, pp. 1641–1654, Dec. 1998.

[28] Z. Liu et al., "GeoRGS: Geometric regularization for real-time novel view synthesis from sparse inputs," IEEE Trans. Circuits Syst. Video Technol., vol. 34, no. 12, pp. 13113–13126, Dec. 2024.

[29] W. Wang et al., "ZPressor: Bottleneck-aware compression for scalable feed-forward 3DGS," in Adv. Neural Inf. Process. Syst. (NeurIPS), 2025.

[30] ISO/IEC JTC1/SC29/WG04, "Common test conditions for MPEG immersive video," MPEG document N00659, 2025.

[31] A. Wieckowski et al., "VVenC: An open and optimized VVC encoder implementation," in Proc. IEEE Int. Conf. Multimedia Expo Workshops (ICMEW), Jul. 2021.

[32] A. Dziembowski, D. Mieloch, J. Stankowski, and A. Grzelka, "IV-PSNR—The objective quality metric for immersive video applications," IEEE Trans. Circuits Syst. Video Technol., vol. 32, no. 11, pp. 7575–7591, Nov. 2022.

[33] A. Dziembowski, W. Nowak, and J. Stankowski, "IV-SSIM—The structural similarity metric for immersive video," Appl. Sci., vol. 14, no. 16, p. 7090, Aug. 2024.

[34] W. Zhang et al., "D-FCGS: Feedforward compression of dynamic Gaussian splatting for free-viewpoint videos," arXiv:2507.05859, 2025.